%% file: m33_CPROPS.tex
\documentclass[traditabstract]{aa} 
\usepackage{graphicx} 
\usepackage{setspace}
\usepackage[switch,pagewise]{lineno} 
\usepackage{rotating} 
\usepackage{natbib} 
\usepackage[usenames]{color} 
\usepackage{multirow}
\usepackage{url} 
\usepackage{ulem} 
\usepackage{supertabular}
\usepackage{xcolor}

\usepackage{afterpage} 
\bibpunct{(}{)}{;}{a}{}{,} 
\newcommand{\ratio} {N({\rm H}_2) / I_{\rm CO(1-0)}} 
\newcommand{\ratioo} {N({\rm H}_2) / I_{\rm CO}}

\newcommand{\kms} {{\rm ~ km \ s^{-1}}}

\newcommand{\Xunit} {~{\rm cm^{-2}/(K\kms)}} 
\newcommand{\mum} {\ensuremath{\rm  ~ \mu m} }
 
\newcommand{\dix}[1] {\ensuremath{10^{#1}}}
\newcommand{\xdix}[1] {\ensuremath{\times10^{#1}}}

\newcommand{\HI}{\ion{H}{i}}
\newcommand{\pg}[1]{{  #1}}

\usepackage{txfonts}

\begin{document}
%\linenumbers
%
\title{Giant molecular clouds in the Local Group galaxy M33}
%   \title{Molecular Clouds in the outer disk of M33}
\titlerunning{GMCs in the Local Group galaxy M33} \subtitle{}
\author{P.~Gratier \inst{1,2} \and J.~Braine \inst{1} \and N.J.~Rodriguez-Fernandez \inst{2} 
\and K.F.~Schuster \inst{2} \and C.~Kramer \inst{3} 
% \and E.~M.~Xilouris \inst{4} \and F.~S.~Tabatabaei \inst{5} \and C.~Henkel \inst{5} 
\and E.~Corbelli \inst{4} 
%\and F.~Israel \inst{7} 
\and F.~Combes \inst{6} 
%\and T.~Wiklind \inst{11}
\and N.~Brouillet \inst{1}  
\and P.~P. van~der~Werf \inst{5} 
%\and D.~Calzetti \inst{8} \and S.~Garcia-Burillo \inst{9} \and A.~Sievers \inst{3} 
%\and F.~Herpin \inst{1} \and S.~Bontemps \inst{1} \and S.~Aalto \inst{12} \and B.~Koribalski \inst{13} \and F.~van~der~Tak \inst{14} \and M.~C.~Wiedner \inst{10} 
\and M.~R\"ollig \inst{7} 
%\and B.~Mookerjea \inst{8}
 }

\institute{Laboratoire d'Astrophysique de Bordeaux, Universit\'e de Bordeaux, OASU, CNRS/INSU, 33271 Floirac, France\\
\email{gratier@obs.u-bordeaux1.fr} 
\and IRAM, 300 Rue de la piscine, F-38406 St Martin d'H\`eres, France 
\and Instituto Radioastronomia Milimetrica (IRAM),  Av. Divina Pastora 7, Nucleo Central, E-18012 Granada, Spain
%\and Institute of Astronomy \& Astrophysics, National Observatory of Athens, I. Metaxa \& V. Pavlou, P. Penteli,15236 Athens, Greece 
%\and Max-Plank-Institut f\"ur Radioastronomy (MPIfR), Auf dem H\"ugel 69, 53121 Bonn, Germany 
\and INAF - Osservatorio Astrofisico di Arcetri, L.E. Fermi, 5 - 50125 Firenze, Italy 
\and Leiden Observatory, Leiden University, PO Box 9513,  NL 2300 RA Leiden, The Netherlands 
%\and Department of Astronomy - LGRT, University of Massachusetts - Amherst, 710 North Pleasant Street, Amherst, MA 01002, USA 
%\and Observatorio Astronomico Nacional (OAN)-Observatorio de Madrid, Alfonso XII, 3, 28014-Madrid, Spain 
\and Observatoire de Paris, LERMA, CNRS, 61 Av de l'Observatoire, 75014 Paris 
%\and Space Telescope Science Institute, 3700 San Martin Drive, Baltimore, MD 21218, USA 
%\and Department of Radio and Space Science, Chalmers University of Technology, Onsala Observatory, SE-439 94 Onsala, Sweden 
%\and CSIRO Astronomy \& Space Science, Australia Telescope National Facility, P.O. Box 76,  Epping NSW 1710, Australia  
%\and SRON Netherlands Institute for Space Research, Landleven 12, 9747 AD Groningen, The Netherlands 
\and 1.Physikalisches Institut, Universit\"at zu K\"oln, Z\"ulpicher Str. 77, D-50937 K\"oln, Germany 
%\and Department  of Astronomy \& Astrophysics, Tata Institute of Fundamental Research, Homi Bhabha Road, Mumbai 400005, India
}

\date{}
\abstract{We present an analysis of a systematic CO(2-1) survey at 12" resolution covering most of the Local Group spiral M33, which, at a distance of 840~kpc, is close enough for individual giant molecular clouds (GMCs) to be identified. The goal of this work is to study the properties of the GMCs in this subsolar metallicity galaxy.  The CPROPS (Cloud PROPertieS) algorithm was used to identify 337 GMCs in M33, the largest sample to date for an external galaxy.  The sample is used to study the GMC luminosity function, or mass spectrum under the assumption of a constant $\ratioo$ ratio.  We find that $n(L)dL \propto L^{-2.0\pm0.1}$ for the entire sample.  However, when the sample is divided into inner and outer disk samples, the exponent changes from $1.6\pm0.2$ in the center 2~kpc to $2.3\pm0.2$ for galactocentric distances larger than 2~kpc. On the basis of the emission in the FUV, H$\alpha$, 8$\mu$m, and 24$\mu$m bands, each cloud was classified in terms of its star-forming activity -- no star formation or either embedded or exposed star formation (visible in FUV and H$\alpha$).  At least one sixth of the clouds had no (massive) star formation, suggesting that the average time required for star formation to start is about one sixth of the total time for which the object is identifiable as a GMC.  The clouds without star formation have significantly lower CO luminosities than those with star formation, whether embedded or exposed, a result that is  presumably related to the lack of heating sources.  Taking the cloud sample as a whole, the main non-trivial correlation is the decrease in cloud CO brightness (or luminosity) with galactocentric radius.  The complete cloud catalog, including the CO and HI spectra and the CO contours overlaid on  the FUV, H$\alpha$, 8$\mu$m, and 24$\mu$m images is presented in the appendix.}


\maketitle

\section{Introduction}
At 840 kpc, M33 is the closest spiral in which cloud positions can be unambiguously defined.
M33 is considerably ``younger'' than the Milky Way in that its gas fraction is higher, the stellar 
colors are bluer, and there has been less chemical enrichment by means of nucleosynthesis.
The oxygen abundance in M33 is half that of the Galaxy but there is much local variation and a weak gradient.
\citet{Gratier.2010a}, hereafter Paper I, presented sensitive and high-resolution observations of 
the CO(2--1) and \ion{H}{i} 21~cm lines at a resolution of 12$''$ or 48 pc at the distance of M33.
The CO observations cover about half the area of M33 as a broad strip roughly along the 
major axis and extending out to the edge of the optical disk, just beyond $R_{25}$.
The \ion{H}{i} cube covers the entire disk.  

Molecular clouds are believed to form from atomic 
gas and then, at some point, to produce stars.  The details of this cycle 
HI $\rightarrow$ H$_2$ $\rightarrow$ stars, and in particular the triggering of the first phase change, 
remain uncertain, although several mechanisms are currently invoked 
\citep{Blitz.2006, Krumholz.2008, Gnedin.2009}.  There is also some evidence that the star 
formation process may be somewhat more efficient in this environment \citep{Gardan.2007,Leroy.2006,Braine.2010,Gratier.2010a}.
The galaxy M33 is sufficiently close  that we can compare with Galactic molecular clouds in order to understand 
these differences. In this paper, we describe the identification of a large sample of molecular 
clouds and compare the properties of these clouds with those of Galactic molecular clouds.

The Milky Way molecular cloud spectrum appears dominated by massive clouds 
\citep[$n(m) \propto m^{-1.6}$][hereafter SRBY]{Solomon.1987}
but this may not be true for M33.  \citet{Engargiola.2003} derived a mass spectrum of
$n(m) \propto m^{-2.6}$ from their interferometric CO(1--0) observations, which, at face value, 
would imply that most of the molecular mass in M33 is in the form of small clouds, 
below their sensitivity limit.  \citet{Rosolowsky.2007a} used the same BIMA observations combined 
with NRO 45meter and the 14m FCRAO data from \citet{Heyer.2004} and found that $n(m) \propto m^{-2.0}$.
The observations presented in Paper I are of much higher sensitivity at similar or higher 
resolution and are thus ideal for the construction of a mass spectrum.

While our goal is to understand the mass spectrum of molecular clouds, owing to the uncertainty in the
$\ratioo$ factor we hereafter discuss the CO luminosity function.  Since earlier work in the Galaxy 
or in external galaxies uses a constant $\ratioo$ factor, the functions are directly comparable.
A large sample is necessary to fit a mass function \citep{Maschberger.2009}; we identified 337 molecular clouds
in M33, which is the largest sample beyond the Galaxy to date.

The CO luminosity function of clouds is important but it is equally important to explore the properties
related to the star formation of the clouds.  We are fortunate that M33 has been so thoroughly observed 
at other wavelengths that we can straightforwardly search for star formation related to the clouds.
The lifecycle of a molecular cloud includes at least the following phases: pre-star formation, embedded
star formation, and exposed star formation.  The last phase corresponds in principle to clouds where the 
massive stars have pierced the molecular cloud, allowing H$\alpha$ and/or UV radiation to escape.
The first phase should not contain any warm dust, as traced by the mid-IR radiation at 8 and 24$\mu$m.
The intermediate phase should have warm dust and PAHs heated by the young stars whose optical 
and UV emission is absorbed by the surrounding molecular gas.  On the basis of 8$\mu$m \citep{Verley.2009} and
 24$\mu$m \citep{Tabatabaei.2007} data from Spitzer, in addition to the H$\alpha$ \citep{Greenawalt.1998,Hoopes.2001}
and the FUV emission from the GALEX satellite \citep{Thilker.2005}, we classify the clouds as phase A, B, or C
with A being the first phase  described above.
%
%In the following Section, we describe the cloud finding algorithm \citep[CPROPS][]{Rosolowsky.2006} 
%and how average cloud properties vary with radius.  As part of the investigation of their properties, we 
%compare the cloud size-linewidth relation to the Galactic \citet{Larson.1981} relation.  The cloud luminosity function
%is then calculated for the whole galaxy and for different radii.  Finally, the cloud classification scheme is explained 
%and the figures on which the classifications were based are published as an annex.  From this, we derive 
%relative lifetimes in each phase.
%Our 12$\arcsec$ resolution (48pc) observations are of sufficient resolution to identify individual GMCs.
%In principle, the detection limit is very low (3rms$\times$chanwidth$\times$beamsize).
%Goal is to study the population of clouds in M33 over the full range of observations available 
%and how these properties vary with galactocentric distance.
%%CO and \ion{H}{i} observations are from \cite{Gratier10} and  
%After briefly describing the observations, the cloud-finding algorithm is presented and the cloud sample  defined.
%
%%In order to study the GMCs in M33 in an unbiased way, the CPROPS

\section{Data}
This study is based on CO(2--1) observations of M33 by the IRAM 30m telescope. This is an ongoing project and this paper focuses on a subset of these data over an area of 650 arcmin$^{2}$ aligned mainly along the major axis of the galaxy. The dataset and data reduction are presented in detail in Paper I. The angular resolution is $12\arcsec$, which corresponds to 48~pc at the distance of M33, and the sensitivity is 20--50 mK in main beam temperature, which is our adopted temperature scale throughout this paper. The atomic gas data is taken from Paper I with an angular resolution identical to the CO data.

\section{\label{sec.CPROPS}Cloud  {properties}}
A modified version of {\tt CPROPS} (Cloud PROPertieS) \citep{Rosolowsky.2006} was used to identify and measure the GMC properties in the $12\arcsec$ datacube. The {\tt CPROPS} program first assigns contiguous regions of the datacube to individual clouds and then computes the cloud properties from the identified emission. The modifications were made to the second step. CPROPS requires clouds to be at least twice the size of the telescope beam.
However, even with this requirement, one of the dimensions is not necessarily resolved, resulting in an undefined cloud radius.  The use of the bootstrap method was extended to the estimation of both the errors and the median values of the radius and luminosities of each of the clouds. This decreases from more than 100 to 29 the number of clouds whose deconvolved radius was not defined because they were  marginally resolved. The original {\tt CPROPS} measures the linewidths by computing second moments of the cloud spectra. We found that this leads to uncertainties greater than when Gaussian  profiles were fitted to the spectra, \pg{as  found by \citet{Gratier.2010} in NGC 6822, where the same 
solution was adopted.} A further modification to CPROPS was thus to measure line widths via Gaussian fitting of the cloud-averaged spectra.

The bootstrapping method consists in drawing a large number of random samples from the initial distribution, allowing the same data to be drawn more than once.  For example, the uncertainty in a quantity derived for a cloud containing 500 $(x,y,v,t)$ pixels can be estimated by drawing 500 pixels randomly from the set.  If each pixel is chosen exactly once, then we have the initial (observed) cloud property.  Since this is a rare occurrence, the greater the variation within the pixels, the greater the resulting uncertainty calculated with the bootstrapping method.  Each time the 500 pixels are drawn yields a value and the distribution of these values yields the uncertainty.  We typically drew the random samples 5000 times.

As an example, we  measured the systemic velocity of a cloud containing N pixels.  To form a ``virtual cloud'', N sample pixels were chosen from the real cloud, allowing the same values to be chosen more than once.  When a pixel was chosen, it had a position, a velocity, and a temperature.  The N (say, 500) pixels in the real cloud can be given numbers from 1 to 500 (N).  The virtual cloud was created by randomly selecting a pixel among those 500 a total of 500 times, such that the same value could be chosen several times.  This set of values could be used to calculate the same quantities (size, linewidth, luminosity, etc) as in the case of the 500 original pixels.  Since the virtual cloud could have the same pixel more than once and also have holes, the values had to be considered as a vector, a series of 500 $(x,y,v,t)$ values that could be used to calculate properties, rather than a physical cloud that of course could not have two pixels identical in both  position and velocity.

This step was performed K (usually 5000) times.  Each of the K bootstrap samples could be written as a vector $[(x_{i},y_{i},v_{i},T_{i})_{i\in(1,N)}]_{k\in(1,K)}$.
For each of these samples, the first moment along the velocity axis was computed to be
\begin{equation}
V_{k} = \frac{\sum_{i=1}^N v_{i} T_{i}}{\sum_{i=1}^N T_{i}}
\end{equation}
Since this was repeated 5000 (K) times with different samples chosen from among the same cloud's pixels, it is 
straightforward to obtain a distribution.
The value and uncertainty in the cloud's systemic velocity was then taken as, respectively, the median and rms-dispersion of the $V_{k}$ distribution.

\begin{figure}
	[htbp]
	\begin{center}
	  \includegraphics[angle=0,width=8.8cm]{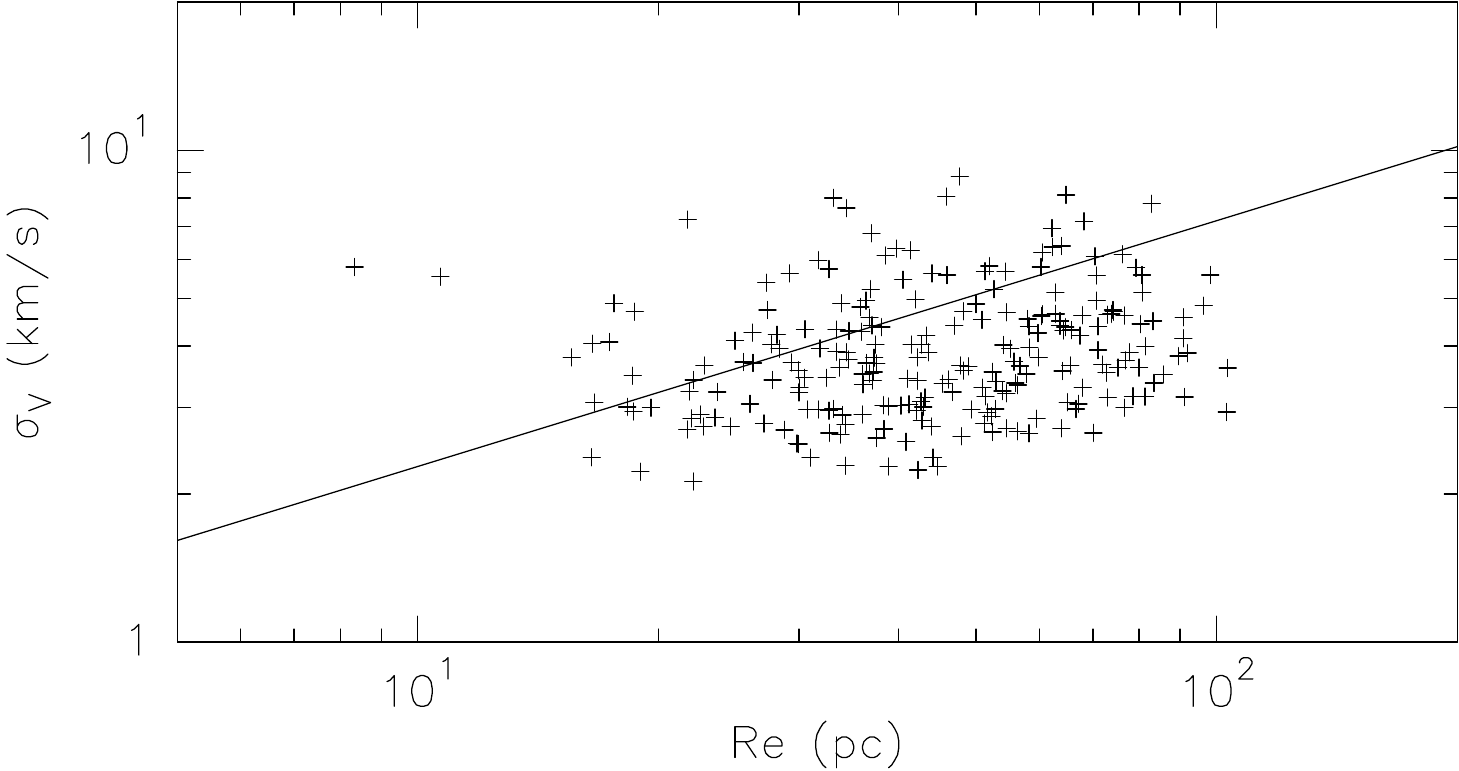}
		\caption{\label{fig.larson} {Size-linewidth plot of M33 clouds.  Crosses indicate the velocity dispersion
of the M33 clouds as a function of the deconvolved radius $R_{e}$  calculated by CPROPS (cf. Sect. 3.1 \citet{Rosolowsky.2006}).
Note that to obtain the full velocity width at half maximum, the dispersion needs 
to be multiplied by $2\sqrt{2\ln2}\simeq 2.35$.
The line represents the scaling law found by SRBY in the Milky
Way.}}  
	\end{center}
\end{figure}

 {Figure~\ref{fig.larson} shows the scaling law between the size and the linewidth for the M33 clouds. No apparent correlation is visible between the size and linewidth and there is a large scatter in the linewidths.  The bulk of the points lie below the scaling law found by SRBY in the Milky Way. \pg{The use of second moments to measure linewidths provides very similar results but with more scatter.  We note that \citet{Blitz.2007} found similar results for Local Group galaxies and the outer Galaxy.} This means that the clouds in M33 are either larger for a given linewidth or have a smaller linewidth for a given size.}

 {One reason why the clouds do not follow the SRBY and  \cite{Larson.1981} relationships between linewidth and size might be that the dynamic range in size is too small because our observations, at $\sim 50$ pc resolution, were unable to resolve all the clouds and measure their size. In the SRBY sample, the largest clouds had sizes comparable to the ones we found in M33 (100pc), but the smallest were ten times smaller than the clouds we observed in M33.} 

%The $\sim 50$ pc resolution of the CO data is not sufficient to measure cloud sizes.  This is presumably 
%why the cloud sample does not follow the \cite{Larson.1981} relationship between size and linewidth.

%Explain why not using Mvir\\
%Explain why gaussian fits to lines\\
%Explain why median and IQR for HI\\
%Explain catalog images (cf online data)\\
Table~\ref{tab.clouds} summarizes the properties of the 337 GMC we identify in M33.
 The columns indicate the cloud ID, the signal-to-noise ratio, the intensity-weighted central position expressed in arcseconds as an offset with respect to the central position in both RA and DEC,
the deprojected distance from the center of M33, the effective radius of the cloud (designed to be equivalent to the SRBY definition), the velocity and linewidth of the GMC \pg{(determined by a Gaussian fit to the line profile with no attempt to correct for finite channel width)}, the FWHM of the HI emission within the sky projected cloud contours, and the molecular and atomic gas masses (not including helium) within the CO cloud contours.  The H$_2$ mass is calculated from the CO luminosity following the prescription in Paper I of  using a $\ratio$ factor of $4\times\dix{20}\Xunit$ which is  twice the usually assumed ``Galactic'' conversion factor \citep{Dickman.1986} and a {CO(2--1)/CO(1--0)} ratio equal to 0.73.  {The atomic hydrogen luminosities of the clouds were computed by summing the observed \ion{H}{i} intensity over the same projected area as the CO cloud and over the whole velocity range.} This is the largest sample of molecular clouds identified in an external galaxy.  As we demonstrate, sample size is very important in deriving reliable statistical measurements.

We verified that all 337 clouds are indeed real detections (see figures in appendix).
The CPROPS parameters were chosen to be fairly conservative to avoid spurious false positives
and the varying noise level over the region observed was input to CPROPS to avoid finding clouds  regions affected by high levels of noise.

 {In the rest of the paper, when average values of a parameter are given, they are computed as a weighted mean using the inverse of the square of the uncertainty as weights.  Thus, poorly defined values cannot lead not incorrect results.}

\section{GMC luminosity function}
%The shape of the cloud luminosity function gives an information on the relative proportion of small and large giant molecular clouds. 
Previous studies of giant molecular clouds luminosity functions in both the Galaxy and  Local Group galaxies \citep[e.g.][]{Williams.1997,Rosolowsky.2005,Rosolowsky.2007a} have shown that a truncated power-law
\begin{equation}
\label{eq.TPareto}
\frac{dN}{dL}\propto \left(\frac{L}{L_{max}}\right)^{-\alpha}
\end{equation}
is adequate to describe the observed luminosity function of GMCs.
However, the estimation of the power-law parameters (exponent $\alpha$ and truncation value $L_{max}$) can be strongly biased. 

\subsection{Method}
\citet{Maschberger.2009} extensively tested different estimation methods (linear fitting of the histogram, fitting of the cumulative distribution function, and the maximum likelihood method) and concluded that the maximum likelihood method based on work by \citet{Aban.2006} is both numerically stable and unbiased for a large number of clouds.
The estimated truncated power-law cumulative distribution is described by 
\begin{equation}
\label{eq.cumuest}
P(l>L) = \frac{L^{1-\widehat{\alpha}}-\widehat{L}_{\min}^{1-\widehat{\alpha}}}{\widehat{L}_{\max}^{1-\widehat{\alpha}}-\widehat{L}_{\min}^{1-\widehat{\alpha}}},
\end{equation} 
where $\widehat{L}_{\min}$ is the estimated lowest luminosity, $\widehat{L}_{\max}$ the estimated truncation value, and $\widehat{\alpha}$ the estimated exponent. 
We refer to \citet{Maschberger.2009} for details of the computation of the estimated values of the parameters.

We determined a completeness limit of our sample by computing the luminosity of the the smallest cloud that the \texttt{CPROPS} algorithm can identify. This hypothetical cloud has an area twice the beam area of our observations and an integrated intensity equal to that of a Gaussian function with a three channel FWHM and a peak intensity four times the noise level.
\begin{equation}
\label{eq.completeness}
L_{min}\simeq 4\times2\sigma\times\Delta_{v}\times 2\times \Omega_{lobe}^2 \simeq 6\times10^3~{\rm K\,\kms pc^2},
\end{equation}
for a characteristic noise level of $\sigma \simeq 50$~mK, a  $\Delta_{v}=2.6\kms$ channel width, and a $\Omega_{lobe}^2= 2700~{\rm pc^2}$ beam  area. 
The $4 \times 2\sigma$ factor comes from a single channel at 4$\sigma$ with a channel at 2$\sigma$ intensity on either side such that the sum is $4 \times 2\sigma$.
Owing to local variations of the map noise level, the completeness limit can vary by 
$\sim 30$\% from cloud to cloud (see Fig. 3 in Paper I).  In the following, the power-law parameter estimation is made over the fraction of clouds that have a luminosity larger than $8\times10^3~{\rm K\,\kms pc^2}$.

The large number of clouds in our sample allows an exploration of the variation in the power-law parameters as a function of radius in M33. We divided our cloud sample into two and three subsamples containing an equal number of clouds. The first and second columns of Table~\ref{tab.expo_result} lists, respectively, the number and fraction of total clouds above this limit for the different radii bins used.
It is immediately apparent that dividing the sample into three (equal) subsamples results in large uncertainties for each of the subsamples, illustrating the importance of sample size.

\subsection{Uncertainties}
To estimate uncertainties in both the power-law exponent and truncation value, we used a bootstrapping method. 
In the case of cloud luminosities, 164 luminosities were drawn from the 164 values in Tab. A1 that are above our completeness limit, allowing the same value to be drawn more than once.  Each set of 164 values was then used to estimate $L_{max}$ and $\alpha$ ($L_{min}$ is held at $8 \times 10^3$ km/s/pc2).  The process of drawing values and calculating Lmax and $\alpha$ is repeated 5000 times yielding the set of values in the inset of Figure 2 (only $\alpha$ is shown, Lmax is roughly the mass of the largest cloud in the sample).  The median and the dispersion of this histogram are then used to determine Lmax and $\alpha$ and their uncertainties (Table 2).
 {

We varied two parameters between the CPROPS  runs, namely the threshold value to include emission (from 1.5$\sigma$ to 2.5$\sigma$), and the minimum area for a region to be considered as a cloud (from one to two times the beam area).
Varying these options led to changes in the luminosity function estimated parameters that are much smaller than the estimated uncertainties. This is because modifying CPROPS input values almost exclusively influences  the number of faint clouds well below the completeness limit we have used determine the luminosity function.}

\subsection{The GMC CO luminosity function in M33}
\begin{figure*}
	[p]
	\begin{flushleft}
	  \includegraphics[angle=0,width=8.8cm]{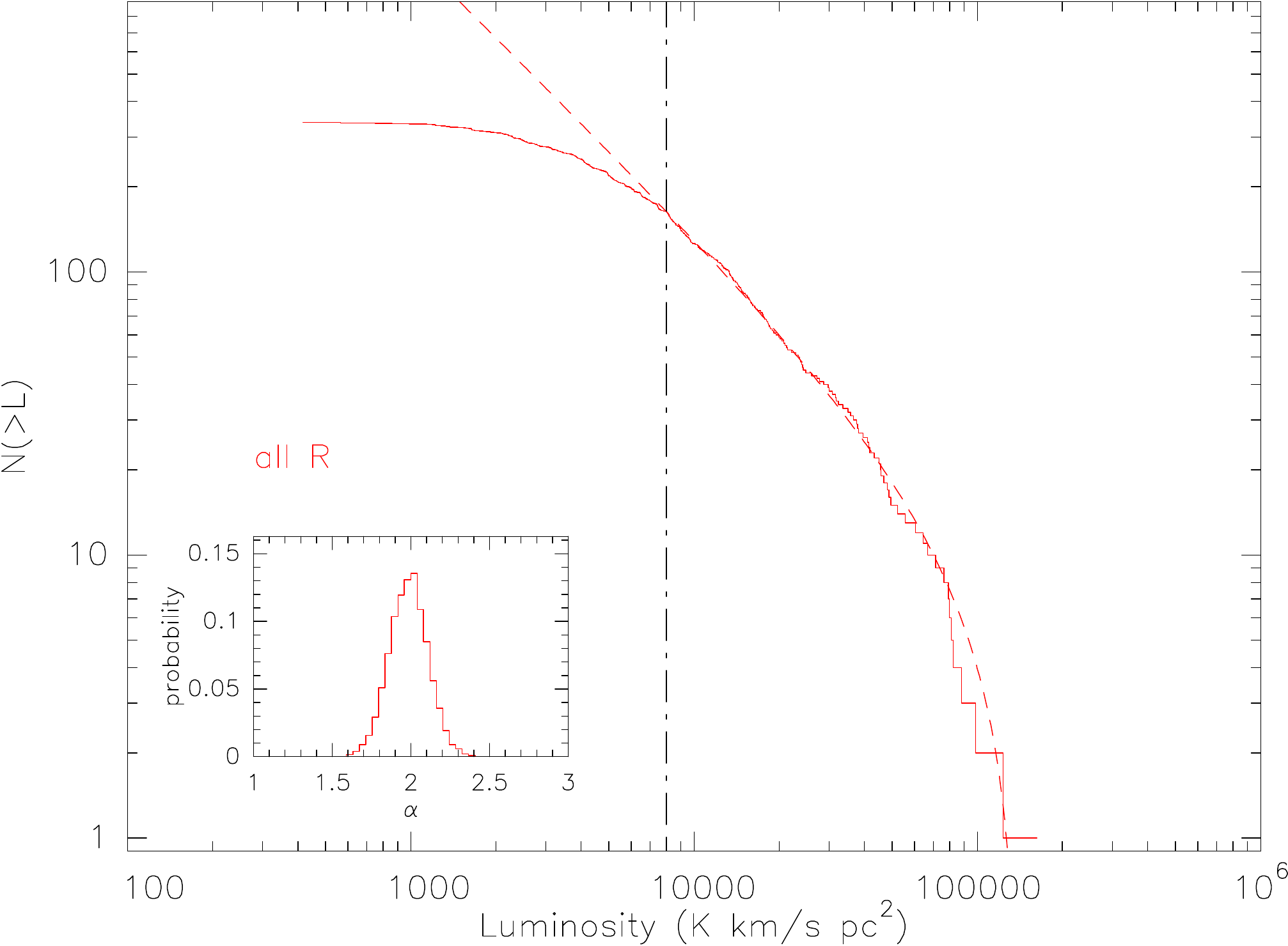}
		\includegraphics[angle=0,width=8.8cm]{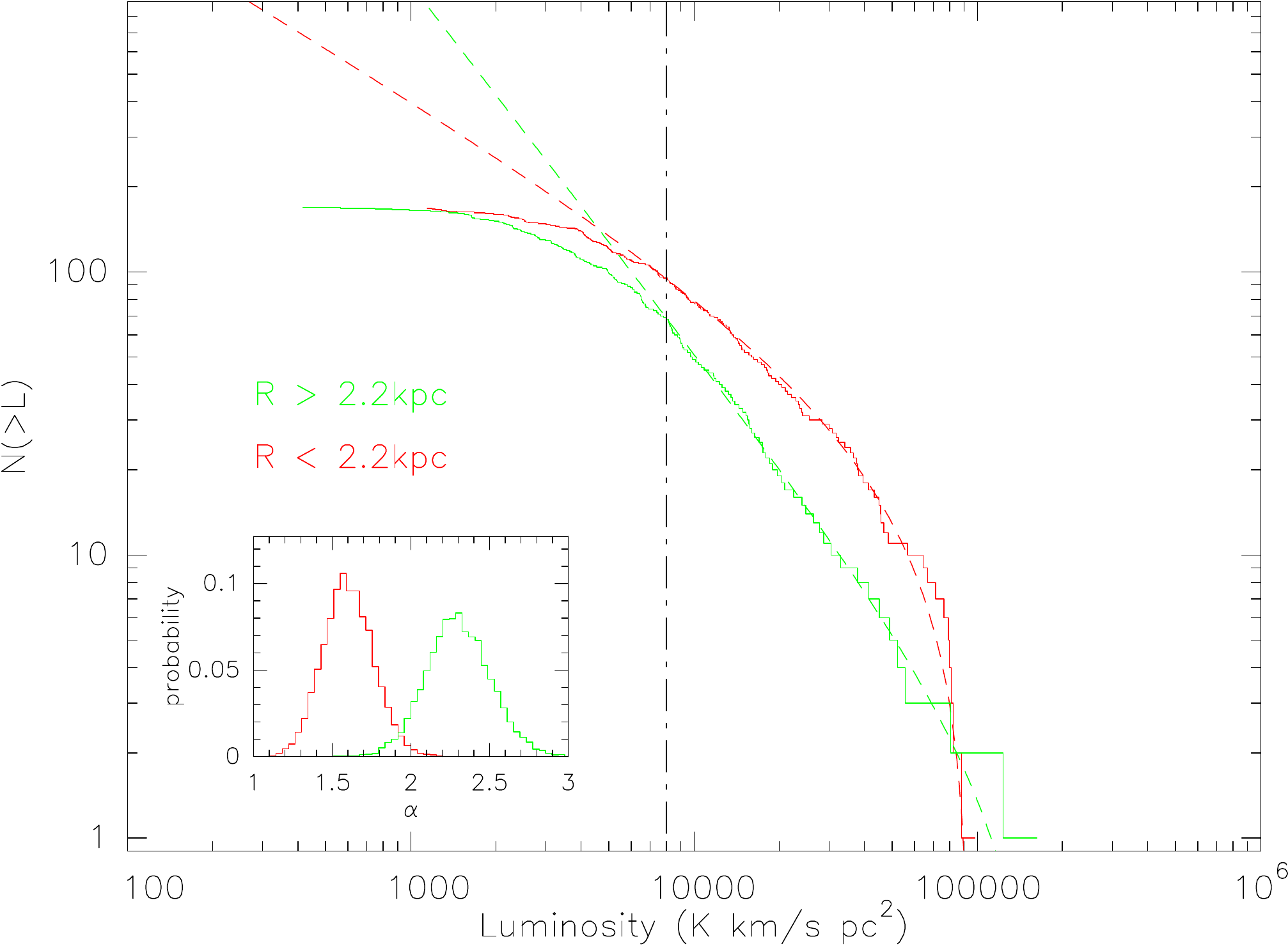}  
		\includegraphics[angle=0,width=8.8cm]{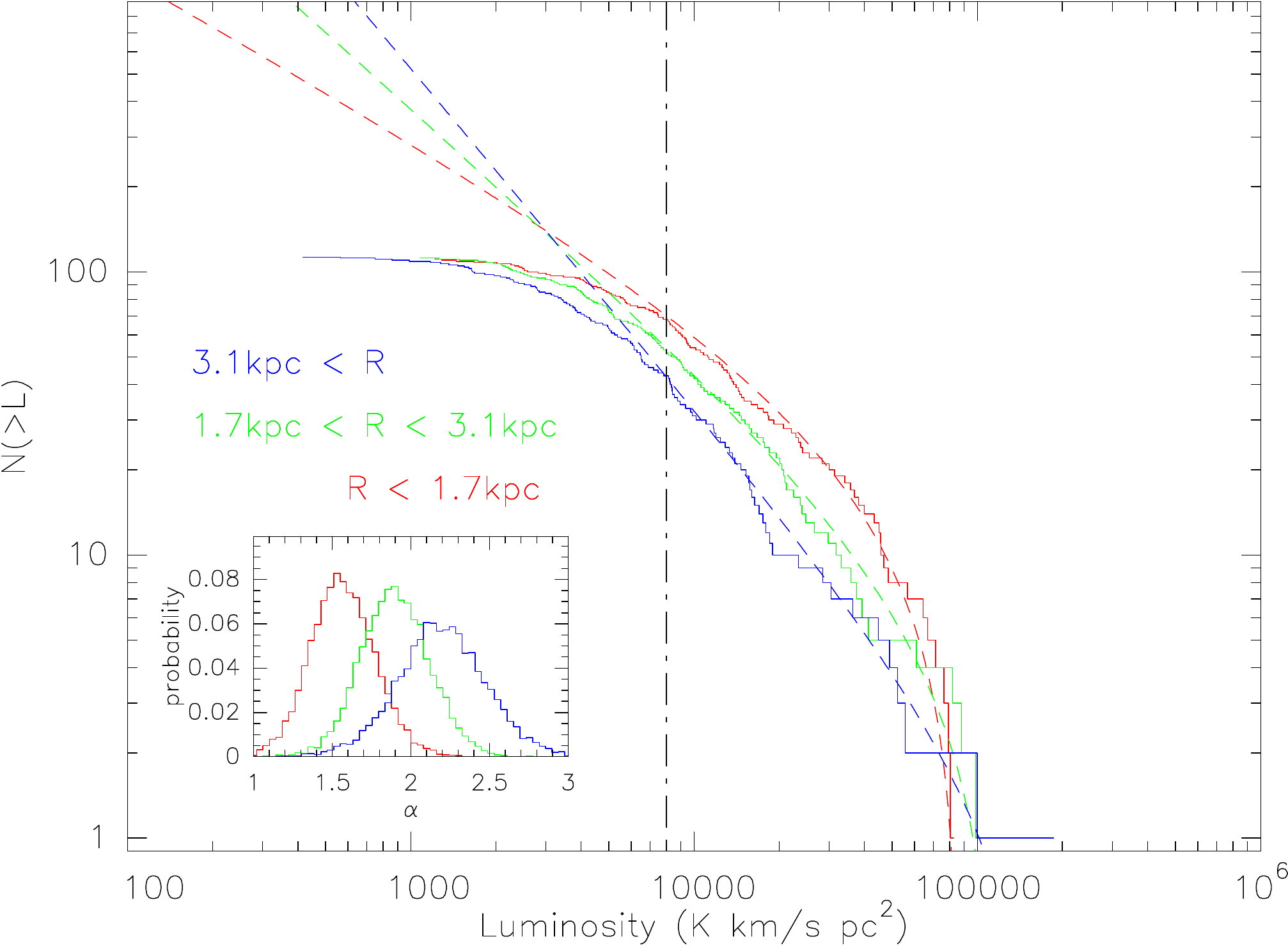} 
		\caption{\label{fig.masshisto}Cumulative luminosity distributions for the GMCs in M33.  The upper left panel includes clouds at all radii in a single luminosity function; the upper right panel divides the cloud sample into two equal samples separating the inner disk clouds  from those at larger radii.  The lower panel divides the sample into three parts (see Tab~\ref{tab.expo_result} for numerical results). In each panel, the range in radii for each curve is indicated following the color coding of the curves.  The solid curves represent the real data, whereas the dashed lines represent the luminosity function calculated from maximum likelihood estimation of the power-law exponent and truncation luminosity.
The insets to the lower left of each panel indicate the distribution in slope values found with the bootstrapping method.
While there is substantial overlap when three radial bins are used (lower panel), showing that randomly
selected samples can yield somewhat different results, the distributions are quite separate when only two 
radial bins are used.  This is reflected in the uncertainties.
The vertical lines indicate the completeness limit; only these clouds are used to estimate the luminosity function parameters.}  
	\end{flushleft}
\end{figure*}

The three panels of Fig.~\ref{fig.masshisto} show both the observed and modeled cumulative luminosity functions for the entire cloud set and for two radial binnings. Table~\ref{tab.expo_result} summarize the parameter values and uncertainties for these radial binnings. \pg{The first column is the range in radii considered, the second column is the number of clouds above the completeness limit, the third column is the corresponding fraction of the total number of clouds, and the two last colums are, respectively, the power-law exponent and power-law truncation luminosity.} In each case, the luminosity function is correctly modeled by a truncated power-law. The exponent for the sample of clouds spanning the entire M33 disk is $2.0\pm0.1$ similar to the value found by \citet{Rosolowsky.2007a}.  Looking at how the luminosity function varies with radius, we find that the exponent increases between the central and the external parts of M33. The relative importance of less luminous clouds  thus increases going towards the exterior of M33.  {\citet{Rosolowsky.2007a} found the same trend but only at the $1-1.5\sigma$ level (inner and outer indices of $\alpha = -1.8 \pm 0.2$ and $-2.1 \pm 0.1$). For the truncation mass of the power-law, we replicate the results of \citet{Rosolowsky.2007a} finding that the truncation is more pronounced in the inner galaxy than in the outer regions. }.  The $1.6\pm0.2$ value for clouds within the central 2~kpc is very similar to the one found for Galactic GMCs \citep{Solomon.1987,Rosolowsky.2005}. The external value is greater than 2, implying that the majority of the cloud luminosity is found in small clouds. \citet{Rosolowsky.2005} found a similar steepening of the power-law in Milky Way clouds but this is the first time that this trend is  {clearly} observed in an external galaxy.

Fixing the completeness limit ($L_{min}$), and only estimating $L_{max}$ and $\alpha$ using the values above that limit does not change their values as long as $L_{min}$ is near the real completeness limit.  If $L_{min}$ is set to zero or a very small value, then the global fit to the data is very poor.

\begin{figure*}[p]
	\centering
		\includegraphics[angle=270,width=88mm]{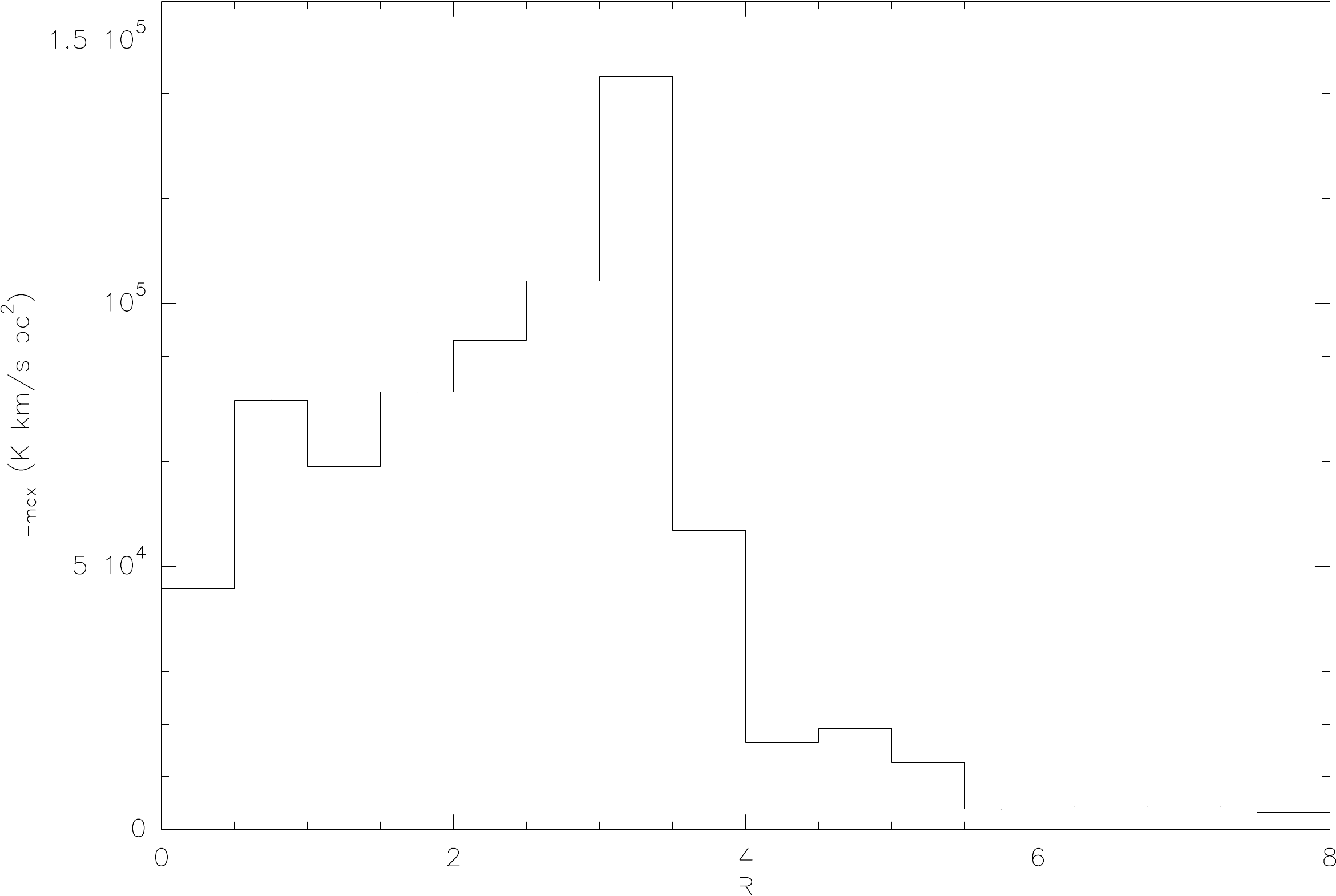}
	\caption{ Peak CO luminosity of the GMCs as a function of radius.  Note the absence of bright GMCs beyond $\mathrm{R=4~kpc}$.}
	\label{fig.fig_Lmax_rad}
\end{figure*}

Assuming that the CO luminosity function represents the cloud mass function, the change in the exponent $\alpha$ means that the molecular gas mass goes from being dominated by large clouds in the inner part to smaller clouds beyond 2.1 kpc.  This is not seen in the $L_{max}$ value, which remains at a value of about $10^5~{\rm K\,\kms pc^2}$ because large clouds are present until about $R \sim 3.5$ kpc and the bin corresponding to radii larger than 3.1~kpc still contains some very luminous clouds (NGC 604 in particular).  As shown in Fig.~\ref{fig.fig_Lmax_rad}, the cloud luminosity then drops precipitously such that only small clouds are present beyond 4 kpc.  Unfortunately, the number of clouds beyond 4 kpc is too small to enable us to calculate either a mass or luminosity function, despite the large area mapped beyond this galactocentric distance.  The calculation of these functions is one of the goals of our ongoing completion of the survey of M33.

 { What is the physical meaning behind the steepening of the GMC luminosity function with radius?
One of the most obvious possibilities is that we  simply detect the effect of a decreasing $\ratioo$ such that clouds near the center have stronger CO emission per H$_2$ molecule.  This could be due to ($i$) a metallicity gradient,  ($ii$) a temperature gradient that reduces the surface brightness of molecular clouds in CO, or  ($iii$) a density gradient that could make collisional excitation of CO less efficient at large radii \citep[e.g. as seen in NGC 4414 by ][]{Braine.1997} and thus reduce the CO brightness of clouds.  All three are likely to contribute because a metallicity gradient is present, the dust temperature and radiation field decrease with radius, and we have no reason to believe that NGC 4414 is an exception in this respect.

\pg{If $\ratioo$ increases with radius, then for a given cloud mass
the CO luminosity will tend to be lower at large galactocentric radius.
However, since $\ratioo$ is only meaningful for a sample, it cannot be used to 
deduce the mass of an individual cloud (which explains why it is not used 
to deduce the masses of individual Galactic clouds).  As such, this will 
increase the number of low-luminosity clouds in the outer bin.  However,
given that there are some very CO luminous clouds beyond 2.1 kpc 
(e.g. NGC 604), the increase in $\ratioo$ does not simply shift the cloud luminosity
spectrum towards lower luminosities but steepens the slope instead, 
since the high-luminosity points remain present.}

The other possibility is that we detect a true decrease in molecular cloud mass.  It is quite easy to imagine that the radial decrease in gas surface density could result in slower cloud assembly such that star formation stops cloud growth before outer disk clouds reach the masses of inner disk clouds.  We still do not know what triggers molecular cloud formation but in general the processes suggested become less efficient at larger radii.
If pressure is the main driver of the \HI\ to H$_2$ process \citep{Elmegreen.1994,Blitz.2006}, then H$_2$ formation should indeed decrease rapidly towards the outer disk.  If H$_2$ originates mainly from merging \HI\ clouds 
\citep{Brouillet.1992, Ballesteros-Paredes.1999, Hennebelle.1999, Heitsch.2005},
the situation is less clear; \HI\ line widths may be larger in the inner disk but seem to reach a constant level in the outer disk \citep{Dickey.1990,Petric.2007}.  \citet{Krumholz.2008} suggested that a combination of column density and metallicity determines the molecular fraction -- again, this leads to a lower average H$_2$ fraction further out in galactic disks.  We are not yet in a position to distinguish between these possibilities but the on-going observations of M33 are designed with this goal in mind and provide information about a subsolar metallicity environment where molecular gas is expected to require more shielding owing to the smaller dust content. }

\begin{table}[htbp]
\caption[]{\label{tab.expo_result}Results for the luminosity function parameters computed from the maximum likelihood method, uncertainties are from bootstrapping.}
	\begin{center}
		\begin{tabular*}{88mm}{@{\extracolsep{\fill}}lrrrr} \hline\hline\noalign{\smallskip}
 Radii &N\tablefootmark{a}& Fraction& $\widehat{\alpha}$ & $\widehat{L}_{\max}$ \\
 (kpc) &&&  & $({\rm K\,\kms pc^{-2})}$ \\ 
\noalign{\smallskip}\hline\noalign{\smallskip}
%all & 164 & 0.48& {1.98} {+-0.126} &  {16} {+-2.7e4}\\
%\hline
%R{2.2}& 95&0.56&  {1.64} {+-0.17} &  {9.8} {+-0.6e4}\\
%R{2.2}& 69&0.40& {2.29} {+-0.21} &  {19} {+-5.7e4}\\
%\hline
%R{1.7}& 68&0.60& {1.63} {+-0.22} &  {8.7} {+-0.3e4}\\
% {1.7}R{3.1}& 52& 0.46& {1.89} {+-0.22} &  {12} {+-1.5e4}\\
% {3.1}R& 44&0.39& {2.22} {+-0.29} &  {21} {+-9.2e4}\\
 all & 164 & 0.48& $2.0\pm0.1$ &  $16\pm2.7\xdix{4}$\\
\hline
R$<${2.2}& 95&0.56&  $1.6\pm0.2$ &  $9.8\pm0.6\xdix{4}$\\
R$>${2.2}& 69&0.40& $2.3\pm0.2$ &  $19\pm5.7\xdix{4}$\\
\hline
R$<${1.7}& 68&0.60& $1.6\pm0.2$ &  $8.7\pm0.3\xdix{4}$\\
 {1.7}$<$R$<${3.1}& 52& 0.46& $1.9\pm0.2$ &  $12\pm1.5\xdix{4}$\\
 {3.1}$>$R& 44&0.39& $2.2\pm0.3$ &  $21\pm9.4\xdix{4}$\\
 \hline
		\end{tabular*}
		\tablefoot{
		\tablefoottext{a}{Number of clouds above the $L_{compl}=8\times10^3 {\rm K\,\kms pc^{2}}$ completeness limit. See Fig.~\ref{fig.masshisto} for the corresponding luminosity functions and comparison with data.}}
	\end{center}
\end{table}

\section{Cloud Types}
In this section we present classification of the clouds as a function of their star forming properties.
\subsection{Defining GMC types}

As we are interested in how clouds evolve into stars, the sample clouds were classified 
as either those without detected star formation (A), those with embedded star formation (B), or those with exposed star
formation (C).  The idea is that this should represent an evolutionary sequence with class C 
preceding cloud dispersal.  The relative fraction of each cloud type should then represent 
the relative lifetime of that phase.  \cite{Kawamura.2009} did a similar classification of the clouds in the 
Large Magellanic Cloud. 

The clouds were defined in the CO datacube in such a way that the dispersed phase is not present.
The clouds were classified based on the figures presented in Appendix~\ref{app.booklet}.  The
8\mum\ and 24\mum\ images were used to trace embedded star formation and the H$\alpha$ and GALEX FUV 
emission to determine whether the star formation is visible.  A major problem is that although M~33 is 
a disk galaxy, it is not always clear in crowded regions whether the continuum emission in any of these bands
is actually associated with the CO cloud or not.  This is a particular problem for the H$\alpha$ and FUV
where large regions are seen in emission and it is sometimes difficult to attribute the H$\alpha$ or FUV emission
to the object generating the CO and IR emission.  Six testers among the co-authors classified the clouds independently.  While for some clouds the situation is quite clear, for others
there was substantial dispersion among the ``testers". In each figure, the cloud classifications are given with the proportion of the different types found by the tester.

 {The criteria for a given classification were deliberately left 
without any strict flux thresholds.
The general idea --- no visible star formation in A clouds, embedded 
(i.e. 8 and 24 micron emission but not seen in H$\alpha$ or FUV) star 
formation in B clouds, and the C classification when the cloud was 
considered as detected in all bands --- is clear.  However, applying
these criteria is in practice not so trivial.  For example, in a crowded 
field, should one associate emission in a given band that is not 
centered on the cloud with the molecular cloud? Different testers 
evaluated this differently and this provides a measure of the uncertainty.
Outer disk clouds 
are in general weaker than inner disk clouds, hence a threshold flux 
would  be inappropriate.  Furthermore, the threshold, which is 
necessarily a somewhat arbitrary value, has a major effect in determining 
cloud classifications.  In the work by \citet{Kawamura.2009}, slightly varying 
the H$\alpha$ flux level has a strong impact on the cloud 
classifications.  Our clouds have a single-peak H$\alpha$ flux distribution,
thus we could completely determine our B versus C classification by changing
the flux level at which a B cloud becomes a C cloud.  Furthermore,
we used two fluxes for each part of the classification (8 and 24, 
FUV and H$\alpha$), unlike the \citet{Kawamura.2009} classification. 
In the end, we decided that the classification should be made by the observer's 
eye, particularly when emission was present that was not centered on clouds. 
The full catalogue is available so readers can actually do their 
own classification.}

\subsection{Proportion of each type of GMC}
\begin{figure}
	[p]
	\begin{center}
		\includegraphics[angle=270,width=8.8cm]{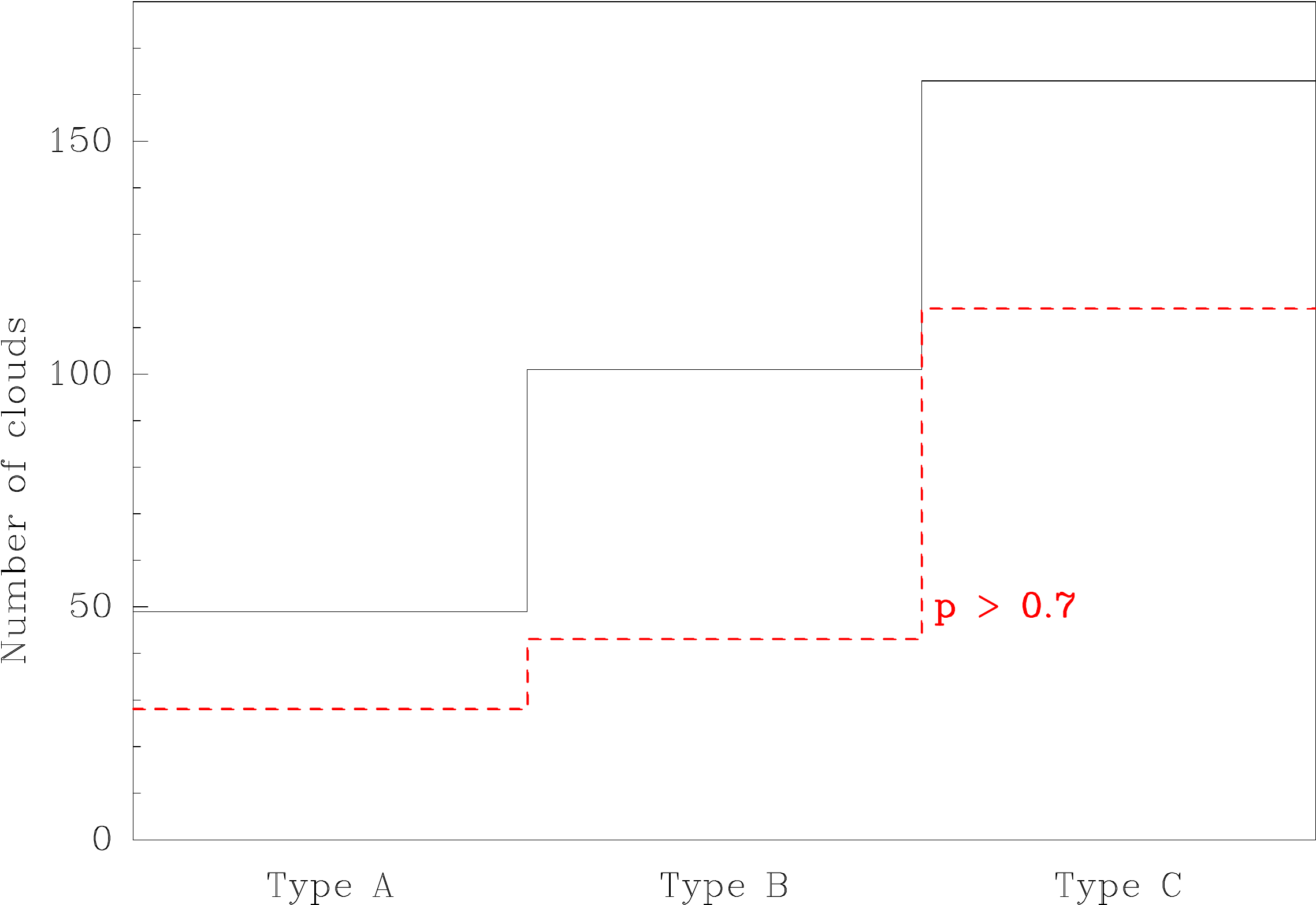}
		\caption{\label{fig.CldFrac}Histogram of the
		proportions of the different cloud types in our
		catalog. Type~A are GMCs without detected star
		formation, type~B GMCs with embedded star formation
		and type~C GMCs with exposed star formation. The black line corresponds to all the clouds with an unambiguous identification, and the red dashed line to the clouds where at least four out of six tester agreed on the cloud type.}
	\end{center}
\end{figure}
\begin{table}[htbp]
\caption{\label{tab.CldFrac}Proportions of the different cloud types in
our catalog.}
	\begin{center}
		\begin{tabular*}{88mm}{@{\extracolsep{\fill}}lrrrr} \hline\hline\noalign{\smallskip}
 &Type A\tablefootmark{a}& Type B\tablefootmark{b}& Type C\tablefootmark{c} & Other \\
 \noalign{\smallskip}\hline\noalign{\smallskip}
 Average (\%)&17.0 & 32.6&48.3&2.1\\
 Dispersion (\%) &4.3&17.1 &19.3&4.1\\ 
  \noalign{\smallskip}\hline\noalign{\smallskip}
		\end{tabular*}
		\tablefoot{
		\tablefoottext{a}{GMCs without detected star formation}
		\tablefoottext{b}{GMCs with embedded star formation}
		\tablefoottext{c}{GMCs with exposed star formation}
		}
	\end{center}
\end{table}
By having several ``testers'' identify each cloud's type independently, we were able to determine both an average and a dispersion for each of the different types. Table~\ref{tab.CldFrac} shows the average and standard deviation of the tester's classifications. The computed dispersion is a measure of the uncertainty in the cloud classification.  
Only type A clouds are defined by an absence of emission in all wavebands considered. They are thus more easily agreed upon between the different observers and this is reflected in the smaller uncertainties for this type compared to the other two (B and C).  
Figure~\ref{fig.CldFrac} shows the number of each type of cloud in the whole sample and in the $p\ge0.7$ subsample (see next subsection).  The sum is less than 337 because 24 clouds had ambiguous identifications -- they were typically classed by an equal number of testers as B and C because the A classifications were more agreed upon.
Type B and C cloud fractions are associated with larger uncertainties as the difference between embedded and visible star formation is somewhat more sensitive to the testers than the absence of star formation. 

The result is that about one sixth of the identified molecular clouds have no associated star formation, while five sixths are associated with star formation. For this subset, it seems that slightly more than half of the clouds are linked to exposed star formation.  A large part of the dispersion was determined by  whether testers associated emission along the line of sight to a cloud, or part of one, as emission associated with the cloud.  When the emission at 8 or 24$\mu$m or in H$\alpha$ or FUV was found to peak at the cloud center as defined by the CO, it was naturally identified as being associated with the cloud. However, when the star formation tracer peaked significantly away from the CO peak, testers had quite different opinions about whether they should attribute the emission to star formation within the GMC.  Since there are likely cases where the star formation is {\it not} associated with the GMC but simply occurs along the same line of sight, it is very likely that some C clouds should in fact be B and that some B should be A.  Velocity-resolved measurements in the star formation tracers, similar to those that could be provided by H$\alpha$ Fabry-Perot measurements, would be of use to estimate the fraction of misclassifications.  This would of course increase the number of A clouds, which we take as a lower limit.

\citet{Kawamura.2009}  made a similar classification of the molecular clouds that they identify in the LMC into three types corresponding to our classification. They used H$\alpha$ luminosity to classify their clouds, their first type (Type I) corresponding to GMCs without associated H$\alpha$ emission. Their other two cloud types were defined as clouds with H$\alpha$ emission that are respectively below (type II) and above (Type III) an H$\alpha$ luminosity threshold. We initially tried to use a similar classification based on a H$\alpha$ luminosity threshold, but because the distribution of H$\alpha$ luminosities from our sample of GMC's is unimodal (i.e. has a single peak), the threshold value determines the relative proportion of type II to III clouds. This is related to the larger uncertainties we found for our type B and C clouds. 

\subsection{Physical properties as a function of cloud type}
\begin{figure*}
	[p]
	\begin{center}
	    \includegraphics[angle=0,width=0.4\textwidth]{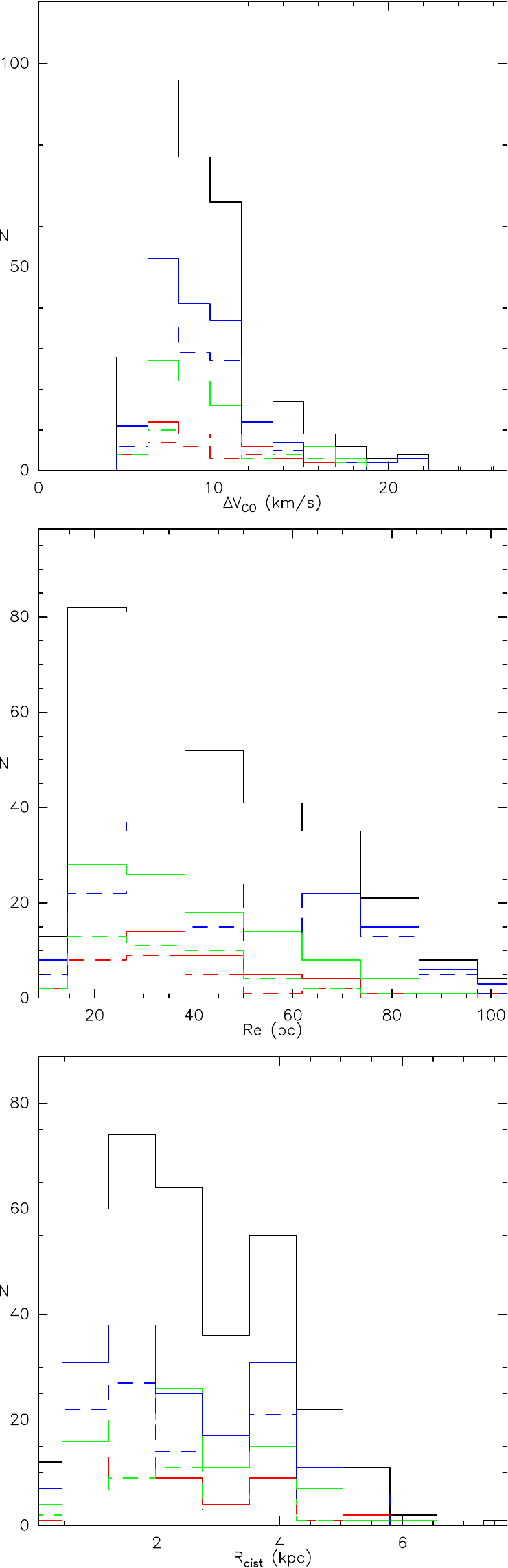}
	    \includegraphics[angle=0,width=0.4\textwidth]{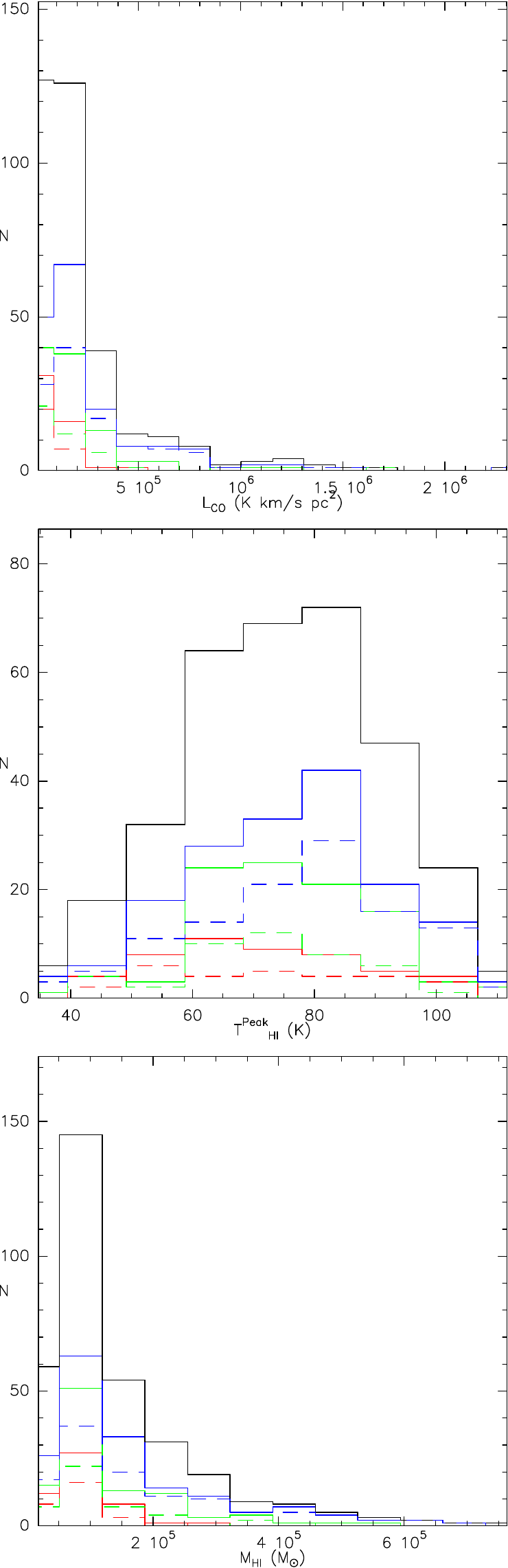}
		\caption{\label{fig.multi_histo}Histograms of the GMC properties. From left to right and top to bottom: CO(2--1) FWHM, molecular gas mass from CO(2--1), effective radius, peak \ion{H}{i} temperature, galactocentric radius, atomic gas mass. In each panel, the thick black line corresponds to the whole cloud sample, the red line to type A clouds, the green line to type B clouds, and the blue line to type C clouds. The solid and dashed versions of each color correspond respectively to no thresholding and a 0.7 probability thresholding for the clouds (see details in text).}  
	\end{center}
\end{figure*}

We wish to determine whether the cloud properties are different for the different cloud types we have identified. In contrast  to the previous paragraph, we here include
%have divided clouds into types by selecting 
only clouds whose types have been identified unambiguously. This gives slightly different proportions of each type as some clouds in our catalog are identified as having equal probabilities of being two different types (e.g. cloud 162 is ambiguously defined). For each property, we have drawn in Fig.~\ref{fig.multi_histo} the histograms of the complete cloud sample, and of the different types. Table~\ref{tab.clt_type} summarizes the property averages and dispersions for each cloud type.
Furthermore, to study the influence of the dispersion introduced by having several testers identify clouds, we give for each property the results for all unambiguously identified clouds and for a subset of these clouds that have been identified as a given type by at least four out of the six testers. These results are indicated with a probability threshold of $p\ge0.7$ in Table~\ref{tab.clt_type}. The uncertainties given are a measure of the dispersion in the property values and are not corrected by a $1/\sqrt{N}$ factor; the uncertainty in the mean is considerably smaller.

The non-star-forming clouds (type A) have on average a lower luminosity than the star-forming (type B or C) clouds but the luminosity is correlated to both the size and  the cloud surface brightness. The effective radius of the type C clouds are larger than the two other types. Nevertheless, the average surface brightness of the non-star-forming clouds is fainter than that of the star-forming GMCs; their lower luminosity is not therefore only due to the type C clouds being larger in size.  A second reason for grouping the B and C clouds together is that they are not always intrinsically different -- a cloud with exposed star formation on one side will be classed as either  B or C depending on the angle of view.

The significance of the differences between the properties of  the star-forming clouds (types B and C) and the non-star-forming clouds (type A) were assessed using a Kolmogorov-Smirnov test by taking into account the uncertainties in the measured properties. The usual K-S test yields a value $p$ of the probability that two samples are drawn from the same distribution. In the case of noisy data, one can simulate a large number of samples of property values taking into account the uncertainties (in our case assuming Gaussian errors and a dispersion as measured by CPROPS as described in Sect.~\ref{sec.CPROPS}) and apply a KS test to each of these samples. The distribution of the $p$ values was then summarized in terms of its mean $\left<p\right>$ and dispersion $\sigma_{p}$. The null hypothesis, {\it i.e.}  that the samples are drawn from the same distribution, is rejected if both $\left<p\right>$ and $\sigma_{p}$ are small.

We applied this method to all measured cloud properties, each time drawing  5000 random samples, and the results for $\left<p\right>$ and $\sigma_{p}$ are shown in Table \ref{tab.KS}.  {Except for the peak CO temperature, the CO luminosity, and the \HI\ luminosity, all results are compatible with the properties of the star-forming clouds being similar to those of the non-star-forming clouds. 
The difference between the luminosity distributions is driven by the absence of luminous molecular or atomic clouds that are not associated with star formation, as can be seen directly in the panel corresponding to the CO and \HI\ luminosities in Fig.~\ref{fig.multi_histo}. Whether this is an effect of the limited spatial coverage of the CO observations will be determined once the full disk has been mapped.}

% {The values of the \ion{H}{i} luminosities are also significantly different but as neither the \ion{H}{i} integrated intensity, nor the peak \ion{H}{i} temperatures differ in the two groups, this result is weaker.}   {As the cloud sizes (Re) are not statistically different for the two cloud groups, the difference in the cloud luminosities must arise from brightness temperature differences as shown by the significant difference between the peak CO temperature distributions of the two cloud groups .  }

\begin{table}[htbp]
\caption{\label{tab.KS}Results of the Monte Carlo Kolmogorov-Smirnov test for the properties of star-forming clouds against non-star-forming clouds.}
\begin{center}
\begin{tabular*}{88mm}{@{\extracolsep{\fill}}lrr}
  \hline
  \hline
  & $<p>$ &$\sigma_{p}$\\
  \hline
  R$_{dist}$ &0.97 & $\ldots$\\
Re & 0.47 &0.29\\
$\sigma_{V_{\element{CO}}}$ & 0.44 & 0.16\\
$\sigma_{V_{\ion{H}{i}}}$& 0.47 & 0.14 \\
L$_{\element{CO}}$ & 0.0021 & 0.0061 \\
L$_{HI}$ & 0.0030 & 0.0033 \\
T$^{peak}_{\element{CO}}$& 0.000077 & 0.00014 \\
T$^{peak}_{\ion{H}{i}}$& 0.32 & 0.25\\
I$_{\element{CO}}$ &0.096& 0.14 \\ 
I$_{\ion{H}{i}}$ & 0.71 & 0.22\\
\hline
\hline
\end{tabular*}
\end{center}
\end{table}

%\begin{table}[htbp]
%\caption{\label{tab.KS}}
%\begin{center}
%\begin{tabular*}{88mm}{@{\extracolsep{\fill}}lrr}
%  \hline
%  \hline
%  & $<p>$ &$\sigma_{p}$\\
%  \hline
%  R$_{dist}$ &0.97 & $\ldots$\\
%Re & 0.47 &0.29 (.45 .26)\\
%$\sigma_{VCO}$ & 0.44 & 0.16 (.43 .11)\\
%$\sigma_{V\ion{H}{i}}$& .47 & .14 (.45 0.1)\\
%L$_{CO}$ & 0.0021 & 0.0061 (0.00036 0.00034) \\
%L$_{HI}$ & 0.0026\\
%T$^{peak}_{\ion{H}{i}}$& 0.12\\
%T$^{peak}_{CO}$& 0.000077 & 0.00014 (0.000032 0.000025)\\
%I$_{CO}$ &0.096& 0.14 (0.035 0.032)\\ 
%I$_{HI}$ & 0.9892066\\
%\hline
%\hline
%\end{tabular*}
%\end{center}
%\end{table}

\begin{table*}[htbp]
\caption{\label{tab.clt_type}Cloud properties as a function of cloud type and for all clouds together.}
	\begin{center}
		\begin{tabular*}{\textwidth}{@{\extracolsep{\fill}}llrrrrrr}
\hline\hline\noalign{\smallskip}
 & &N\tablefootmark{a} & $R_{dist}$ & $R_{e}$ & $ \Delta V_{CO}$&$ \Delta V_{HI}$&M$_{H_{2}}$ \\
 & && (kpc) & (pc) & (km/s)&(km/s) &(M$_{\sun}$) \\
\noalign{\smallskip}\hline\noalign{\smallskip}
\multicolumn{2}{c}{All clouds} & 337	&$2.47\pm1.39$ & $55\pm22$ & $6.5\pm2.0$&$11.1\pm2.7$&$8.5\pm12\times\dix{4}$\\
\noalign{\smallskip}\hline\noalign{\smallskip}
\multirow{2}{*}{Type A}& No filtering& 49	&$2.53\pm1.38$	&$51\pm24$	&$5.5\pm1.9$	&$11.0\pm2.9$	&$3.7\pm5.0\times\dix{4}$	\\
						& $p\ge0.7$		& 28 	&$2.46\pm1.31$	&$49\pm28$	&$7.3\pm2.8$	&$11.4\pm2.9$	&$4.5\pm3.6\times\dix{4}$	\\
\noalign{\smallskip}\hline\noalign{\smallskip}
\multirow{2}{*}{Type B}& No filtering	& 101	&$2.46\pm1.26$	&$40\pm19$	&$7.6\pm2.2$	&$10.9\pm2.3$	&$9.2\pm11\times\dix{4}$	\\
						& $p\ge0.7$		& 43	&$2.42\pm1.28$	&$44\pm16$	&$8.4\pm3.2$	&$11.0\pm2.4$	&$7.2\pm11\times\dix{4}$	\\
\noalign{\smallskip}\hline\noalign{\smallskip}
\multirow{2}{*}{Type C}& No filtering	& 163	&$2.40\pm1.45$	&$59\pm23$	&$6.8\pm2.1$	&$11.2\pm2.8$	&$11\pm14\times\dix{4}$	\\
						& $p\ge0.7$		& 114	&$2.51\pm1.47$	&$60\pm22$	&$7.2\pm2.1$	&$11.0\pm2.5$	&$13\pm16\times\dix{4}$	\\
\noalign{\smallskip}\hline\noalign{\smallskip}
 & &N & M$_{\ion{H}{i}}$  & T$_{\ion{H}{i}}^{peak}$ & T$_{CO}^{peak}$ & $\left<I_{CO}\right>$& $\left<I_{\ion{H}{i}}\right>$ \\
 & && (M$_{\sun}$) & (K) & (mK) & (K$\kms$) & (K$\kms$) \\
\noalign{\smallskip}\hline\noalign{\smallskip}
\multicolumn{2}{c}{All clouds} & 337	&$14\pm12\times\dix{4}$ & $74\pm16$ & $38\pm34$&$1.5\pm1.1$&$1.1\pm0.3\times\dix{3}$\\
\noalign{\smallskip}\hline\noalign{\smallskip}
\multirow{2}{*}{Type A}& No filtering	& 49	&$8.6\pm4.9\times\dix{4}$	&$72\pm15$	&$23\pm20$	&$1.0\pm0.7$	&$1.1\pm0.4\times\dix{3}$	\\
						& $p\ge0.7$		& 28 	&$8.1\pm5.0\times\dix{4}$	&$73\pm16$	&$26\pm23$	&$1.0\pm0.9$	&$1.2\pm0.4\times\dix{3}$	\\
\noalign{\smallskip}\hline\noalign{\smallskip}
\multirow{2}{*}{Type B}& No filtering	& 101	&$13\pm10\times\dix{4}$		&$75\pm15$	&$40\pm32$	&$1.7\pm1.2$	&$1.1\pm0.3\times\dix{3}$	\\
						& $p\ge0.7$		& 43	&$12\pm9.0\times\dix{4}$	&$72\pm14$	&$46\pm41$	&$1.4\pm1.5$	&$1.1\pm0.4\times\dix{3}$	\\
\noalign{\smallskip}\hline\noalign{\smallskip}
\multirow{2}{*}{Type C}& No filtering	& 163	&$17\pm14\times\dix{4}$		&$75\pm16$	&$41\pm35$	&$1.6\pm1.1$	&$1.1\pm0.4\times\dix{3}$	\\
						& $p\ge0.7$		& 114	&$18\pm15\times\dix{4}$		&$77\pm17$	&$46\pm39$	&$1.7\pm1.2$	&$1.1\pm0.3\times\dix{3}$	\\
\noalign{\smallskip}\hline\noalign{\smallskip}
		\end{tabular*}
		\tablefoot{
		\tablefoottext{a}{Number of clouds in each sample or subsample.}
		For each cloud type, the first line gives the average and the dispersion for all clouds of that type and the second line gives the same quantities for the clouds whose type is well-defined. }
	\end{center}
\end{table*}

\section {The influence of environment on cloud properties}

How are the properties of molecular clouds related to the environment in which they formed?
Given our sample of clouds, we decided to search for relations between the properties of our CO clouds (namely size, line width, CO luminosity, CO line area, and peak temperature) and: ($a$) galactocentric distance, which enables us to estimate both the stellar mass density and overall gravitational potential; and ($b$) HI properties (same as for CO except that there is no size).  For all of these pairs of properties, we calculated the Spearman rank coefficient $\rho$ .  As it was impossible to introduce the error bars in the individual values into the calculation of $\rho$, the uncertainty in $\rho$ was calculated by drawing many values for each of
the properties and for each cloud, using the uncertainty in each property and each cloud, to calculate $\rho$ many times.  The uncertainty in $\rho$ was then evaluated to be the dispersion in the values of $\rho$ calculated using the real errors in the quantities.

\begin{figure*}
	[p]
	\begin{flushleft}
		\includegraphics[width=\textwidth,clip]{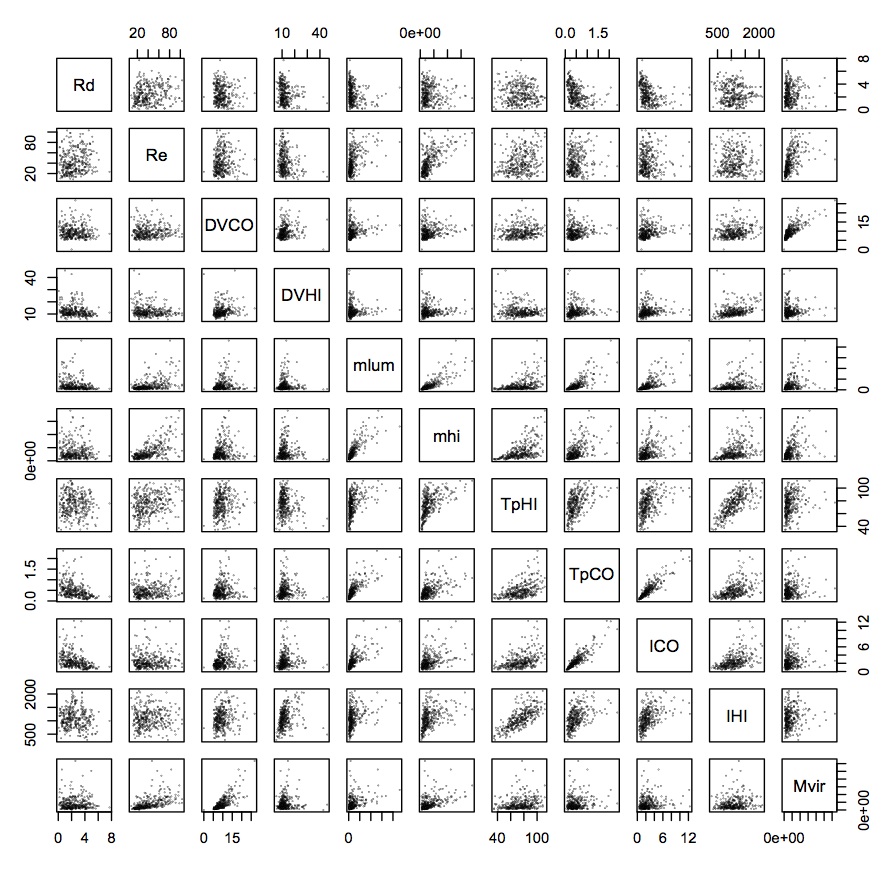} \caption{\label{fig.corr}Plots of all the pair combinations between the cloud properties to show how they correlate.  Each point in each plot represents one of the 337 clouds in the sample.  For each plot, the x-axis corresponds to the parameter on the
diagonal for the same column and the y-axis to the parameter on the diagonal for the same line.  For example, the plot in the eighth column
and fifth row shows the cloud mass as derived from the CO luminosity ("mlum" -- see Table~\ref{tab.clouds}) as a function of the peak CO line temperature ("TpCO") and the scales can be read at the top of the column for the $x-$axis and at the right end of the row for the $y-$axis.
For odd columns and even row numbers, the scales are respectively to the bottom and left.  The units for each quantity are given in Table~\ref{tab.clouds}.}
	\end{flushleft}
\end{figure*}

Table~\ref{tab.corr} contains a square matrix with zeros along the diagonal.  The triangle above the diagonal gives the $\rho$ coefficients between the properties shown as column and row labels.  For example, to find out whether there is a correlation between the CO line width and the HI mass, the coefficient at the intersection of the $\Delta$VCO and mhi lines is $\rho =$0.185 in the upper part.
The triangle below the diagonal gives the dispersion in $\rho$ for the same quantities.  As can be seen in Fig~\ref{fig.corr}, the correlation (if any) is very weak between $\Delta$VCO and mhi, as confirmed by the low correlation coefficient.

 {Many ``trivial'' correlations are found, e.g. between Ico and Tpeak(CO), or between $\Delta$VCO and Mvir ($\Delta$VCO enters strongly into the calculation of Mvir).  Another example of a trivial correlation is between the HI and CO masses (mhi and mlum, respectively) because the radius enters into both as $Re^2$.  Beyond such obviously trivial correlations, the peak CO line temperature and HI mass are correlated, as are the integrated intensities.  While such a correlation is expected on very large scales, such as spiral arms, it is not necessarily expected on the GMC scale investigated here, where much of the HI might have been converted into H$_2$, and it is clearly not present on small scales in the Galaxy.  However, the correlation between CO and HI line widths is considerably weaker, if present at all. }  {As can be seen in the catalog figures (inside the box with the spectra), HI and CO velocities closely agree illustrating that with few exceptions the molecular clouds correspond to \HI\ peaks at the same velocity.  The average difference between the molecular cloud velocity and the HI velocity at the same position is smaller than 2 $\kms$.  Even for a formation time of 10 Myr, this would result in an offset smaller than 20~pc, well within our cloud boundaries.  In their study of outer disk molecular clouds, \citet{Digel.1994} found an average offset of 40~pc between the HI and CO peaks but noted the small dynamic range of the HI column density.  The galaxy M33 has a lower metallicity but similar radiation field as the Milky Way, hence its the molecular clouds can be expected to require more shielding.  We may be seeing this effect here via a closer association.}

The other real correlation is between the galactocentric distance and the peak CO line temperature (or Ico, which is strongly related to Tpeak), CO peak temperatures decreases with galactocentric distance in our sample.   {We note that \citet{Bigiel.2010} with a small sample of clouds at high resolution found the opposite trend.   The steepening of the cloud mass spectrum might generate an effect similar to this if the clouds become more and more diluted in our beam.  However, our interferometric data on distant clouds (in prep) do not support this conclusion and the clouds observed by \citet{Bigiel.2010} may be part of a shell of molecular gas detected by \citet{Gardan.2007a}, that is possibly created (or brightened) by an ejection of matter.}

An interesting absence of or very weak correlation is that between $\Delta$VCO and galactocentric distance.  {One might expect that $\Delta$VCO would decrease with the general level of star-forming activity and rotational shear and indeed in our earlier observations of individual clouds \citep{Braine.2010}, it appeared as though the CO line widths decreased with distance from the center of the galaxy.  Our data do not exclude an (anti)correlation between $\Delta$VCO and galactocentric distance but show that it is weak (at best) at the level of individual clouds, despite the steepening of the luminosity (mass?) function.}

% latex table generated in R 2.7.2 by xtable 1.5-4 package
% Fri Dec 10 10:09:11 2010
\begin{table*}[ht]
\caption{\label{tab.corr}Matrix of the correlation coefficients (upper triangle) and their dispersion (lower triangle) for the cloud properties.}
\begin{center}
\begin{tabular}{rlllllllllll}
  \hline
 & Rd & Re & DVCO & DVHI & mlum & mhi & TpHI & TpCO & ICO & IHI & Mvir \\
  \hline
Rd & 0.000 & 0.111 & -0.101 & -0.176 & -0.206 & -0.003 & -0.017 & -0.441 & -0.323 & -0.054 & 0.026 \\
  Re & 0.042 & 0.000 & 0.069 & -0.075 & 0.255 & 0.355 & 0.122 & 0.065 & 0.038 & 0.016 & 0.281 \\
  DVCO & 0.012 & 0.014 & 0.000 & 0.229 & 0.211 & 0.185 & 0.141 & 0.116 & 0.205 & 0.187 & 0.462 \\
  DVHI & 0.044 & 0.040 & 0.037 & 0.000 & 0.090 & 0.105 & -0.046 & 0.209 & 0.183 & 0.341 & 0.048 \\
  mlum & 0.014 & 0.014 & 0.029 & 0.031 & 0.000 & 0.713 & 0.335 & 0.665 & 0.494 & 0.271 & 0.283 \\
  mhi & 0.046 & 0.016 & 0.012 & 0.036 & 0.043 & 0.000 & 0.475 & 0.533 & 0.352 & 0.427 & 0.346 \\
  TpHI & 0.012 & 0.022 & 0.044 & 0.050 & 0.030 & 0.034 & 0.000 & 0.399 & 0.284 & 0.557 & 0.150 \\
  TpCO & 0.029 & 0.031 & 0.013 & 0.038 & 0.026 & 0.024 & 0.044 & 0.000 & 0.643 & 0.453 & 0.107 \\
  ICO & 0.044 & 0.048 & 0.014 & 0.037 & 0.047 & 0.030 & 0.046 & 0.043 & 0.000 & 0.341 & 0.121 \\
  IHI & 0.027 & 0.032 & 0.027 & 0.041 & 0.025 & 0.023 & 0.046 & 0.038 & 0.043 & 0.000 & 0.099 \\
  Mvir & 0.028 & 0.031 & 0.013 & 0.036 & 0.024 & 0.040 & 0.043 & 0.047 & 0.048 & 0.046 & 0.000 \\
   \hline
\end{tabular}
\tablefoot{As an example, the correlation between galactocentric distance (Rd) and peak CO line temperature is significantly negative, with a correlation coefficient of -0.441 with an uncertainty (dispersion in monte carlo results) of 0.029}
\end{center}
\end{table*}

\section{Comparison with previous studies}
 {Giant molecular clouds in M33 were previously  studied by \citet{Wilson.1990}, \citet{Engargiola.2003}, \citet{Rosolowsky.2003,Rosolowsky.2007a}, and \citet{Bigiel.2010}.  The data presented by \citet{Gratier.2010a} and analyzed here have 12 arcsecond resolution, much higher sensitivities than the other studies, and cover a large fraction of the optical disk.  The linear resolution, while excellent for extragalactic work, is comparable to the size of a GMC so we are interested in comparing with higher resolution studies to understand the effect
of our resolution on the physical properties we derive. 
\citet{Wilson.1990} and \citet{Rosolowsky.2003} observed with a resolution of 20~pc and \citet{Bigiel.2010} with a resolution of 7~pc.  Table~\ref{tab.prev_GMC} describes the subsample of GMCs in our catalog that correspond to GMCs previously identified at higher spatial resolution. These studies were necessarily interferometric and were potentially affected by the lack of short spacings.  
All of the clouds previously detected by \citet{Rosolowsky.2003} and \citet{Bigiel.2010} in the region we mapped in CO(2--1) are present in our catalog. Two categories of clouds can be distinguished, the simple clouds where only one interferometric detected cloud is associated with our own clouds and complex clouds where more than one interferometric cloud corresponds to a given cloud in our sample.

Among the 32 clouds in our sample found in others, 25 remain single clouds at the higher resolution, 5 contain 2 clouds at the higher resolution, and 2 of our GMCs break into 3 at the higher resolution, such that the 32 cloud initial sample actually becomes 41 clouds.  Unsurprisingly, a larger fraction of the multiples are from the Bigiel et al. data.  The positions and velocities are in excellent agreement.  The Bigiel et al. data are at 1.7$"$ resolution and despite the great difference in angular resolution, the cloud contours of our sample follow their clouds extremely well (surrounding them with similar shapes).  The only consistent difference is that the linewidths and cloud sizes derived from the interferometric observations are consistently smaller.
Thus, while the virial masses we derive may sometimes be overestimated owing to the inclusion of more than one cloud in one of our GMCs (owing to our resolution), this appears to be the case for fewer than one quarter of our clouds at 20~pc resolution and only two-fifth when observed at 7~pc resolution.  Even at 1.7$"$ angular resolution, the majority of our clouds appear to remain single clouds.
}

\begin{table}[htbp]
\caption[]{\label{tab.prev_GMC}Table of the correspondance of GMC in our catalog with those identified by  \citet{Rosolowsky.2003} and  \citet{Bigiel.2010}.}
	\begin{center}
		\begin{tabular*}{88mm}{@{\extracolsep{\fill}}lr}
		\hline\hline
		Cloud ID & Corresponding cloud\\
		\hline
		108 & R03--1\\
		95 & R03--2\\
		93/100 & R03--3\\
		125 & R03--4, R03--5, R03--6\\
		92& R03--7\\
		98 & R03--8, R03--9\\
		124 & R03--10, R03--13\\
		120& R03--11\\
		104&R03--12\\
		94&R03--14\\
		103&R03--15\\
		102&R03--16\\
		170&R03--17\\
		196&R03--18\\
		29&R03--19\\
		171&R03--20\\
		25&R03--21\\
		193&R03--22\\
		182&R03--23\\
		256&R03--24,R03--25\\
		242&R03--31\\
		258&R03--33\\
		251&R03--36\\
		245&R03--37\\
		215&R03--40, R03--41\\
		209&R03--42\\
		12&R03--43\\
		\hline
		316&B10-1\\
		286&B10--2\\
		266&B10--3\\
		285&B10-4,B10-5,B10-6\\
		288&B10--7,B10--8\\
		\hline
		\end{tabular*}
		\tablefoot{
		R03 : \citet{Rosolowsky.2003}, B10 : \citet{Bigiel.2010}, Clouds 26,27,28,29,30,39,44 and 45 of B03 are outside of our mapped area
		}
	\end{center}
\end{table}

%\begin{figure*}[p]
%	\centering
%		\includegraphics[angle=270,width=88mm]{figures/Mvirinterf_MlumCO21}
%	\caption{Plot of the Virial masses from interferometric observations against the masses derived from our CO(2--1) luminosities for simple clouds. The clouds present in  \citet{Rosolowsky.2003} are shown as filled squares and the ones in \citet{Bigiel.2010} as open circles.}
%	\label{fig.fig_Mvirinterf_LCO21}
%\end{figure*}

\section{Conclusions}

On the basis of the largest sample  of clouds yet available for an external galaxy, 
our analysis of the cloud population of M~33 has  allowed us to draw 
three sets of conclusions:

Assuming a constant $\ratioo$, the cloud mass spectrum
varies as $n(m) \propto m^{-2.0 \pm 0.1}$ when taken as a whole.
Dividing the sample into two radial bins, the inner disk mass spectrum
follows the proportionality $n(m) \propto m^{-1.6 \pm 0.2}$, while beyond 2.2 kpc the larger number
of less massive clouds steepens the mass spectrum to $n(m) \propto m^{-2.3 \pm 0.2}$.
 {This result was also suggested by \citet{Rosolowsky.2007a} at a low level of significance.}
These exponents are robust to reasonable changes in the completeness limit adopted.
There is a sharp drop in cloud CO luminosity beyond a galactocentric radius of 4kpc
but the number of clouds is insufficient to measure the luminosity (or mass)
function so far from the center.

At least one sixth of the cloud population show no sign of massive  star formation, similar 
that was found for the Large Magellanic Cloud by Kawamura et al. (2009).  These clouds
have lower CO peak temperatures and luminosities, implying that while massive star formation 
is not necessary for detectable CO emission, it increases the CO luminosity.  
Other cloud properties are not statistically significantly different (via the 
Kolmogorov-Smirnov test) between clouds with and without
massive star formation in M~33.

Taking the cloud population as a whole, the average CO luminosity and peak brightness temperature decreases 
with distance from the center of M~33.  Excluding trivial correlations, relations are clearly present  between 
the \ion{H}{i} mass and CO peak temperature and possibly between CO and \ion{H}{i} line widths.
 {On the basis of a more limited sample of 12 clouds, \citet{Braine.2010} found a decrease of the linewidth with galactocentric radius. However for the large sample presented here, any decrease in CO cloud line-width with distance from the center of M33 is not statistically significant according to our criteria}.

\begin{acknowledgements}
We thank the IRAM staff in Granada for their help with the observations. We thank the anonymous referee for its constructive comments and remarks.
\end{acknowledgements}
\bibliographystyle{aa} 
\bibliography{/Users/gratier/Documents/Biblio/biblio}

\begin{appendix}

\onecolumn
\section{GMC Catalog}
\begin{flushleft}
\tablecaption{\label{tab.clouds} Clouds detected in CO.}
\tablefirsthead{
\hline \hline
Cloud & PSNR &$\left.\alpha_{\mathrm{off}}\right.^{\mathrm{a}}$ & $\left.\delta_{\mathrm{off}}\right.^{\mathrm{a}}$ &${\rm R_{dist}}$ &${\rm R_{e}}$ & $\mathrm{V}_{\mathrm{CO}}$ & $\mathrm{FWHM}_{\mathrm{CO}}$ & $\mathrm{FWHM}_{\mathrm{\ion{H}{i}}}$ & $\mathrm{M}_{\mathrm{H_{2}}}$ & $\mathrm{M}_{\mathrm{\ion{H}{i}}}$\\ 
 &  &$(\arcsec)$ & $(\arcsec)$ &(kpc) &(pc) & $(\kms)$ & $(\kms)$ & $(\kms)$ & $\mathrm{M}_{\sun}$ & $\mathrm{M}_{\sun}$\\ 
\hline
}
\tablehead{
\hline
Cloud & PSNR &$\left.\alpha_{\mathrm{off}}\right.^{\mathrm{a}}$ & $\left.\delta_{\mathrm{off}}\right.^{\mathrm{a}}$ &${\rm R_{dist}}$ &${\rm R_{e}}$ & $\mathrm{V}_{\mathrm{CO}}$ & $\mathrm{FWHM}_{\mathrm{CO}}$ & $\mathrm{FWHM}_{\mathrm{\ion{H}{i}}}$ & $\mathrm{M}_{\mathrm{H_{2}}}$ & $\mathrm{M}_{\mathrm{\ion{H}{i}}}$\\ 
 &  &$(\arcsec)$ & $(\arcsec)$ &(kpc) &(pc) & $(\kms)$ & $(\kms)$ & $(\kms)$ & $\mathrm{M}_{\sun}$ & $\mathrm{M}_{\sun}$\\ 
\hline
}
\tabletail{
\hline
\multicolumn{11}{r}{\small\sl continued on following page}\\
}
\tablelasttail{
\hline
\hline
}
\begin{supertabular*}{\textwidth}{@{\extracolsep{\fill}}lrrrrrrrrrr} 
\input{tex_table}
\end{supertabular*} 
\end{flushleft}

\newpage
\section{\label{app.booklet}GMC catalog images}
The following set of 337 figures present for each cloud:
\begin{description}
\item[(\emph{Top left})]
List of cloud properties 
\begin{itemize}
\item GMC number
\item Peak signal-to-noise ratio
\item Position of the GMC in pixels, in relative position with respect to the center of M33, in absolute coordinates
\item CO(2--1) systemic LSR velocity
\item CO(2--1) velocity dispersion
\item GMC effective radius $R_{e}$
\item CO(2--1) luminosity
\item Molecular gas  mass computed from the CO(2--1) luminosity with a $\ratio=4\times\dix{20}\Xunit$ and a constant {CO(2--1)/CO(1--0)} factor of 0.73.
\item Atomic gas mass from the \ion{H}{i} data.
\item Galactocentric distance
\item Histogram showing the distribution of cloud types identified by the ``testers''.
\end{itemize}
\item[(\emph{Top center})]
Spectra of CO(2--1) (\emph{red thick line, left ordinate axis}) and \ion{H}{i} (\emph{thin black line, right ordinate axis}) averaged over the GMC in $T_{mb}$ units. The velocities corresponding to the peak CO(2--1) and \ion{H}{i} emission are also given.
\item[(\emph{Top right})]
Contours of the GMCs in the catalog, the GMC corresponding to the figure is in red. Constant galactocentric radii increasing by 1~kpc are represented by dotted ellipses. The units of the axes are in arcseconds relative to the center of M33.
\item[(\emph{Bottom})]
In these four panels, the CO(2--1) integrated intensity contours for the GMC (\emph{solid white line}, first contour at $80$~mK$\kms$ and following stepped by $330$~mK$\kms$) are plotted on color maps of H$\alpha$, Spitzer 8\mum, GALEX FUV, and Spitzer $24\mum$.
The outer contours of other GMCs in the field (corresponding to the maximum extent of the projection on the plane of the sky of the region identified by  \texttt{CPROPS}) are plotted with a dashed white line. The catalog number of GMCs in the same field of view are given to ease cross-references. The CO(2--1) $12\arcsec$ beam is plotted in the lower left of each subplot.
\end{description}
\begin{figure*}
	[p]
	\begin{flushleft}
		\includegraphics[angle=0,width=17cm]{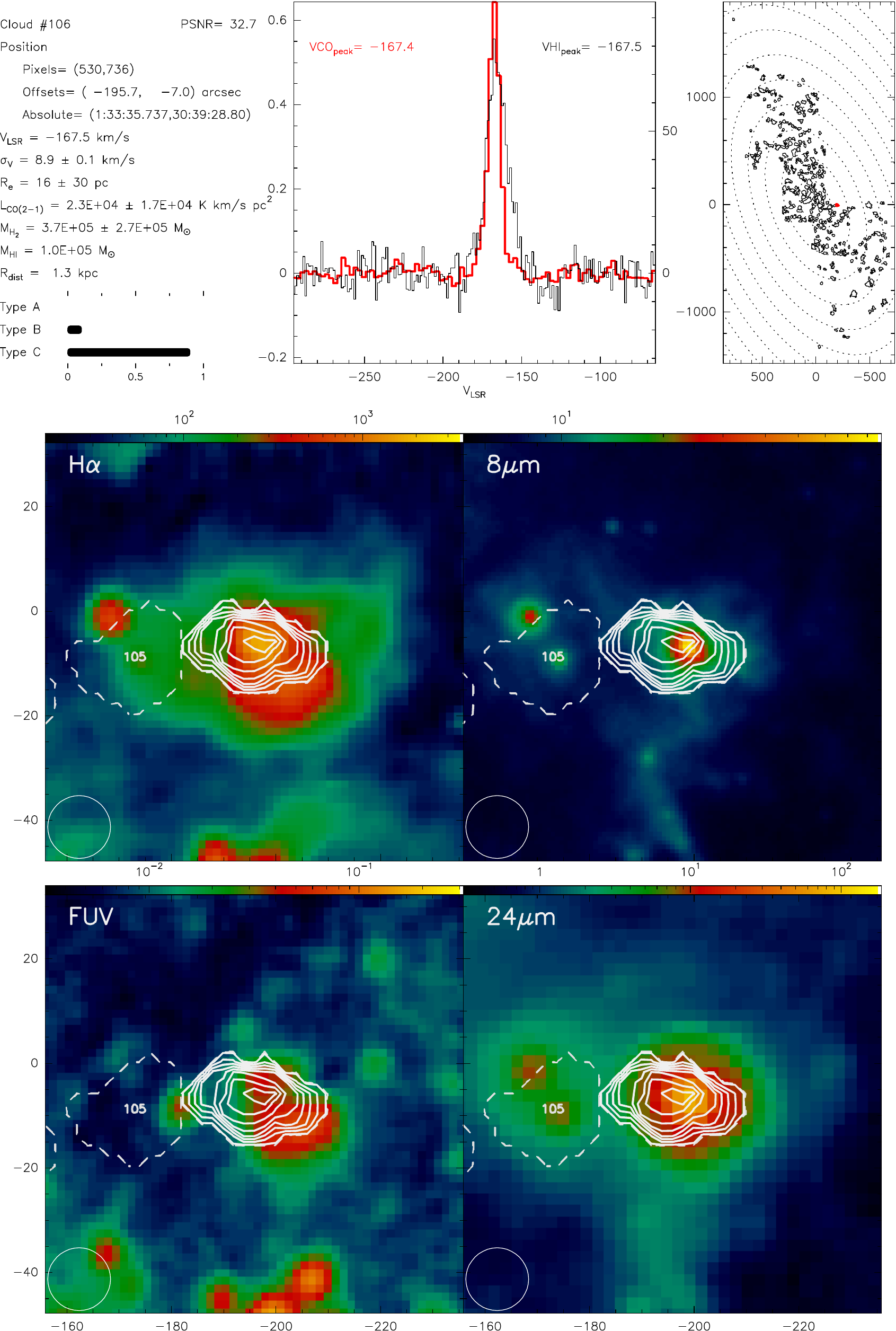} \caption{\label{fig.FIR_CO_radius}Catalogue entry for cloud 106. See description at the beginning of this appendix.}  
	\end{flushleft}
\end{figure*}
\end{appendix}

\end{document}

%% file: tex_table.tex
$1$ & $    9.7$ & $   -503$ & $   -994$ & $    4.6$ & $     32\pm     23$ & $  -97.7$ & $    7.0\pm    0.3$ & $   11.6\pm    0.4$ & $    1.4\pm    0.2\times 10^{5}$ & $    1.3\pm    0.3\times 10^{5}$ \\ 
$2$ & $    7.5$ & $   -239$ & $  -1168$ & $    5.0$ & $     63\pm     18$ & $ -104.9$ & $   12.1\pm    0.5$ & $   12.6\pm    0.7$ & $    2.0\pm    0.4\times 10^{5}$ & $    2.2\pm    0.4\times 10^{5}$ \\ 
$3$ & $    5.1$ & $   -297$ & $  -1236$ & $    5.3$ & $\ldots$ & $ -101.0$ & $    7.6\pm    0.3$ & $   10.6\pm    0.3$ & $    6.1\pm    3.7\times 10^{4}$ & $    9.4\pm    1.9\times 10^{4}$ \\ 
$4$ & $    6.3$ & $   -187$ & $  -1168$ & $    5.1$ & $     20\pm     34$ & $  -98.3$ & $    9.9\pm    0.6$ & $   12.1\pm    0.3$ & $    9.9\pm    3.6\times 10^{4}$ & $    9.4\pm    1.9\times 10^{4}$ \\ 
$5$ & $   10.9$ & $   -159$ & $   -686$ & $    3.0$ & $     70\pm     14$ & $ -102.0$ & $    6.3\pm    0.1$ & $   11.0\pm    0.7$ & $    3.3\pm    0.4\times 10^{5}$ & $    2.0\pm    0.4\times 10^{5}$ \\ 
$6$ & $    6.3$ & $    -83$ & $   -602$ & $    2.6$ & $     23\pm     38$ & $  -97.9$ & $    5.5\pm    0.3$ & $   11.0\pm    0.4$ & $    5.1\pm    3.0\times 10^{4}$ & $    5.0\pm    1.0\times 10^{4}$ \\ 
$7$ & $    4.8$ & $   -268$ & $  -1224$ & $    5.3$ & $\ldots$ & $  -97.1$ & $    8.9\pm    0.0$ & $    9.6\pm    0.4$ & $    3.1\pm    2.2\times 10^{4}$ & $    5.0\pm    1.0\times 10^{4}$ \\ 
$8$ & $    5.2$ & $   -260$ & $  -1231$ & $    5.3$ & $     32\pm     52$ & $ -100.7$ & $    6.0\pm    0.4$ & $    9.6\pm    0.5$ & $    5.6\pm    5.3\times 10^{4}$ & $    5.3\pm    1.1\times 10^{4}$ \\